\pdfoutput=1

\documentclass[twocolumn,twocolumnappendix]{aastex63}
\usepackage{savesym}
\savesymbol{tablenum}
\usepackage{siunitx}
\restoresymbol{SIX}{tablenum}
\usepackage{amsmath,amssymb}

\received{\today}
\revised{\today}
\submitjournal{AJ}
\shorttitle{Synoptic VLBI with the CHIME Pathfinder}
\shortauthors{Leung et al.}
\graphicspath{{./}{figures/}}

\begin{document}

\title{A Synoptic VLBI Technique for Localizing Non-Repeating Fast Radio Bursts with CHIME/FRB}

\correspondingauthor{Calvin Leung}
\email{calvinl@mit.edu}

\author[0000-0002-4209-7408]{Calvin Leung}
\affiliation{MIT Kavli Institute for Astrophysics and Space Research, Massachusetts Institute of Technology, 77 Massachusetts Ave, Cambridge, MA 02139, USA}
\affiliation{Department of Physics, Massachusetts Institute of Technology, 77 Massachusetts Ave, Cambridge, MA 02139, USA}
\author[0000-0002-0772-9326]{Juan Mena-Parra}
\affiliation{MIT Kavli Institute for Astrophysics and Space Research, Massachusetts Institute of Technology, 77 Massachusetts Ave, Cambridge, MA 02139, USA}
\author[0000-0002-4279-6946]{Kiyoshi Masui}
\affiliation{MIT Kavli Institute for Astrophysics and Space Research, Massachusetts Institute of Technology, 77 Massachusetts Ave, Cambridge, MA 02139, USA}
\affiliation{Department of Physics, Massachusetts Institute of Technology, 77 Massachusetts Ave, Cambridge, MA 02139, USA}
\author[0000-0002-3615-3514]{Mohit Bhardwaj}
\affiliation{Department of Physics, McGill University, 3600 rue University, Montr\'eal, QC H3A 2T8, Canada}
\author[0000-0001-8537-9299]{P.J. Boyle}
\affiliation{Department of Physics, McGill University, 3600 rue University, Montr\'eal, QC H3A 2T8, Canada}
\affiliation{McGill Space Institute, McGill University, 3550 rue University, Montr\'eal, QC H3A 2A7, Canada}
\author[0000-0002-1800-8233]{Charanjot Brar}
\affiliation{Department of Physics, McGill University, 3600 rue University, Montr\'eal, QC H3A 2T8, Canada}
\affiliation{McGill Space Institute, McGill University, 3550 rue University, Montr\'eal, QC H3A 2A7, Canada}
\author[0000-0001-6630-7871]{Mathieu Bruneault}
\affiliation{McGill Space Institute, McGill University, 3550 rue University, Montreal, QC H3A 2A7, Canada}
\author[0000-0003-2047-5276]{Tomas Cassanelli}
\affiliation{Dunlap Institute for Astronomy \& Astrophysics, University of Toronto, 50 St. George Street, Toronto, Ontario, Canada M5S 3H4}
\affiliation{David A Dunlap Department of Astronomy \& Astrophysics, 50 St George St, Toronto, Ontario, Canada, M5S 3H4}
\author[0000-0003-2319-9676]{Davor Cubranic}
\affiliation{Department of Physics and Astronomy, University of British Columbia, 325 - 6224 Agricultural Road, Vancouver, BC V6T 1Z1, Canada}
\author[0000-0003-4810-7803]{Jane F. Kaczmarek}
\affiliation{National Research Council Canada, Herzberg Astronomy and Astrophysics Research Centre, Dominion Radio Astrophysical Observatory, PO Box 248, Penticton, British Columbia, V2A 6J9 Canada}
\author[0000-0001-9345-0307]{Victoria Kaspi}
\affiliation{Department of Physics, McGill University, 3600 rue University, Montr\'eal, QC H3A 2T8, Canada}
\affiliation{McGill Space Institute, McGill University, 3550 rue University, Montr\'eal, QC H3A 2A7, Canada}
\author{Tom Landecker}
\affiliation{National Research Council Canada, Herzberg Astronomy and Astrophysics Research Centre, Dominion Radio Astrophysical Observatory, PO Box 248, Penticton, British Columbia, V2A 6J9 Canada}
\author[0000-0002-2551-7554]{Daniele Michilli}
\affiliation{Department of Physics, McGill University, 3600 rue University, Montr\'eal, QC H3A 2T8, Canada}\affiliation{McGill Space Institute, McGill University, 3550 rue University, Montr\'eal, QC H3A 2A7, Canada}
\author[0000-0001-8292-0051]{Nikola Milutinovic}
\affiliation{Department of Physics and Astronomy, University of British Columbia, 325 - 6224 Agricultural Road, Vancouver, BC V6T 1Z1, Canada}
\author{Chitrang Patel}
\affiliation{Dunlap Institute for Astronomy \& Astrophysics, University of Toronto, 50 St. George Street, Toronto, Ontario, Canada M5S 3H4}
\affiliation{Department of Physics, McGill University, 3600 rue University, Montr\'eal, QC H3A 2T8, Canada}
\affiliation{McGill Space Institute, McGill University, 3550 rue University, Montr\'eal, QC H3A 2A7, Canada}
\author[0000-0003-3463-7918]{Andre Renard}
\affiliation{Dunlap Institute for Astronomy \& Astrophysics, University of Toronto, 50 St. George Street, Toronto, Ontario, Canada M5S 3H4}
\author[0000-0001-5504-229X]{Pranav Sanghavi}
\affiliation{CSEE, West Virginia University, Morgantown, WV 26505, USA}
\affiliation{Center for Gravitational Waves and Cosmology, West Virginia University, Morgantown, WV 26505, USA}
\author[0000-0002-7374-7119]{Paul Scholz}
\affiliation{Dunlap Institute for Astronomy \& Astrophysics, University of Toronto, 50 St. George Street, Toronto, ON M5S 3H4, Canada}
\author[0000-0001-9784-8670]{Ingrid H. Stairs}
\affiliation{Department of Physics and Astronomy, University of British Columbia, 325 - 6224 Agricultural Road, Vancouver, BC V6T 1Z1, Canada} 
\author[0000-0003-4535-9378]{Keith Vanderlinde}
\affiliation{Dunlap Institute for Astronomy \& Astrophysics, University of Toronto, 50 St. George Street, Toronto, Ontario, Canada M5S 3H4}
\affiliation{David A Dunlap Department of Astronomy \& Astrophysics, 50 St George St, Toronto, Ontario, Canada, M5S 3H4}

\collaboration{99}{(CHIME/FRB Collaboration)}



\begin{abstract}
    We demonstrate the blind interferometric detection and localization of two fast radio bursts (FRBs) with 2- and 25-arcsecond precision on the 400\,m baseline between the Canadian Hydrogen Intensity Mapping Experiment (CHIME) and the CHIME Pathfinder. In the same spirit as very long baseline interferometry (VLBI), the telescopes were synchronized to separate clocks, and the channelized voltage (herein referred to as "baseband") data were saved to disk with correlation performed offline. The simultaneous wide field of view and high sensitivity required for blind FRB searches implies a high data rate---6.5\,terabits per second (Tb/s) for CHIME and 0.8\, Tb/s for the Pathfinder. Since such high data rates cannot be continuously saved, we buffer data from both telescopes locally in memory for $\approx 40$ s, and write to disk upon receipt of a low-latency trigger from the CHIME Fast Radio Burst Instrument (CHIME/FRB). The $\approx200$ deg$^2$ field of view of the two telescopes allows us to use in-field calibrators to synchronize the two telescopes without needing either separate calibrator observations or an atomic timing standard. In addition to our FRB observations, we analyze bright single pulses from the pulsars B0329+54 and B0355+54 to characterize systematic localization errors. Our results demonstrate the successful implementation of key software, triggering, and calibration challenges for CHIME/FRB Outriggers: cylindrical VLBI outrigger telescopes which, along with the CHIME telescope, will localize thousands of single FRB events to 50 milliarcsecond precision.
\end{abstract}


\section{Introduction} \label{sec:intro}
Fast radio bursts \citep[FRBs,][]{lorimer2007bright,thornton2013population} are brief ($\sim$ ms), usually nonrepeating radio transient events with dispersion measures in excess of that predicted by the electron column density of the Milky Way. Currently, their progenitors and production mechanism are unknown but their high luminosity and impulsive nature have generated significant interest in the astrophysics community \citep{platts2019living}. In addition, due to their cosmological distances~\citep{thornton2013population}, FRB pulses are strongly dispersed by the ionized intergalactic medium and have the potential to probe the large-scale structure of the universe~\citep{mcquinn2013locating,masui2015dispersion,macquart2020census}.

The vast majority of FRBs are not observed to emit multiple bursts~\citep{petroff2016frbcat}\footnote{See \texttt{http://frbcat.org/} for the latest statistics on repeat bursts from known FRB sources.}, and the handful of known repeaters are observed to do so stochastically with the notable exceptions of FRB 180916.J0158+65~\citep{R3} and possibly FRB 121102~\citep{zhang2018fast, rajwade2020possible}. This unpredictability makes localization and followup studies extremely challenging. Since the serendipitous detection of the first FRB in 2007~\citep{lorimer2007bright}, two repeating FRBs have been studied with very long baseline interferometry (VLBI): FRB 121102~\citep{chatterjee2017direct,marcote2017repeating}, with optical followup performed by~\citet{tendulkar2017host}; and FRB 180906.J0158+65~\citep{marcote2020repeating}. The localization of seven others with sufficient precision to identify their respective host galaxies at redshifts between $z = 0.1-0.6$ (180924~\citealt{bannister2019single}, 181112~\citealt{prochaska2019low}, 190523~\citealt{ravi2019fast}, 190102, 190608, 190611, and 190711~\citealt{macquart2020census}) demonstrated a modern-Universe measurement of $\Omega_b$ using FRBs, including the so-called ``missing baryons''. This measurement is consistent with that of \citet{aghanim2018planck}, experimentally evaluating the possibility of using localized FRBs as cosmological probes~\citep{macquart2020census}.

Having detected over seven-hundred FRBs in its first year of operation \citep{RN2}, the Canadian Hydrogen Intensity Mapping Experiment/FRB Project (\citealp{FRBSystemOverview}) has opened up a window for population-level studies of the properties of FRBs~\citep{R1detection,Observation400,RN,RN2,R3,SGR}. However, CHIME/FRB's real-time localization pipeline, which has a precision of arcminutes, does not yet always allow for unambiguous identification of an FRB's host galaxy. For
very bright
FRBs with very low dispersion measure (DM), it is sometimes possible to identify a host by imposing a prior on the host galaxy's maximum redshift.

To routinely pinpoint the host galaxy of FRBs detected by CHIME/FRB, the CHIME/FRB collaboration is developing CHIME/FRB Outriggers, a set of cylindrical telescopes at distances of one hundred to several thousand kilometers from the CHIME telescope. Along with CHIME, the Outriggers will perform a blind VLBI survey to localize thousands of FRBs with 50 milliarcsecond precision. To our knowledge, there has only been one previous attempt to blindly localize FRBs with VLBI. V-FASTR was a campaign to search
for FRBs in archival data taken by the Very Long Baseline Array \citep{wayth2011vfastr, burkespolaor2016limits,wagstaff2016machine}. None was found, highlighting the difficulty of detecting FRBs with traditional radio telescopes. In contrast, the CHIME/FRB Outriggers program will combine CHIME/FRB's high discovery rate with the localization precision afforded by continental baselines, allowing astronomers to conduct detailed population-level studies of FRB host environments.

We report here on the development of a voltage recording backend as a testbed for CHIME/FRB Outriggers that was deployed on the CHIME Pathfinder, itself a reduced-scale testbed for the CHIME telescope~\citep{PathfinderOverview}. We demonstrate a synoptic VLBI calibration technique for CHIME/FRB outriggers, and demonstrate the performance of our technique on automatically triggered single-pulse detections of the bright pulsars B0329+54 and B0355+54. We also localize two FRBs detected during two observing campaigns using CHIME and the Pathfinder in October and December 2019. Our east-west baseline allows for localization of each source in the RA direction on the sky with arcsecond-level statistical uncertainties for bright FRBs.

\section{Instrumentation} \label{sec:instrumentation}
CHIME~\citep{PathfinderOverview} is a beamforming~\citep{CHIMEBeamforming} interferometer located at the Dominion Radio Astrophysical Observatory (DRAO) near Penticton, British Columbia, Canada. It consists of four stationary 20-m $\times$ 100-m parabolic cylindrical reflectors oriented north-south, each of which houses 256 dual-polarization feeds which are uniformly spaced on the focal line of each reflector. Operating as a phased array over the frequency range 400 to 800 MHz, each reflector has a primary beam of 2.6 to 1.3 degrees East-West, directable to any north-south direction from horizon to horizon, with north-south beamwidth increased by the cosecant of zenith angle.

The telescope backend is built with an FX correlator architecture. The first correlator stage, the F-engine, digitizes the analog voltage inputs and spectrally divides the incoming data into 1024 frequency channels over the 400\,$-$\,800~MHz frequency band using a polyphase filter bank~\citep{FEngineOverview}. It is synchronized to a GPS-disciplined ovenized crystal oscillator. The channelized voltage data, hereafter referred to as ``baseband'' data, are passed to the second stage of the correlator
(the X-engine)~\citep{XEngineOverview} at 4 real + 4 imaginary bit depth, for all 1024 frequencies and 2048 signal chains, every $\SI{2.56}{\micro\s}$, for an overall rate of 6.5 Tb/s. In addition to performing real-time processing, the X-engine buffers the baseband data in memory in a 36-s long ring buffer. If the real-time FRB search pipeline~\citep{FRBSystemOverview} detects an FRB candidate, the ring buffer saves the appropriate $\approx\SI{100}{\milli\s}$ segment of data to disk, with the exact duration being determined by the uncertainty in the dispersion measure (DM) estimated by the real-time searc pipeline.

\begin{figure}
    \centering
    \includegraphics[angle=90,origin=c,width=0.45\textwidth]{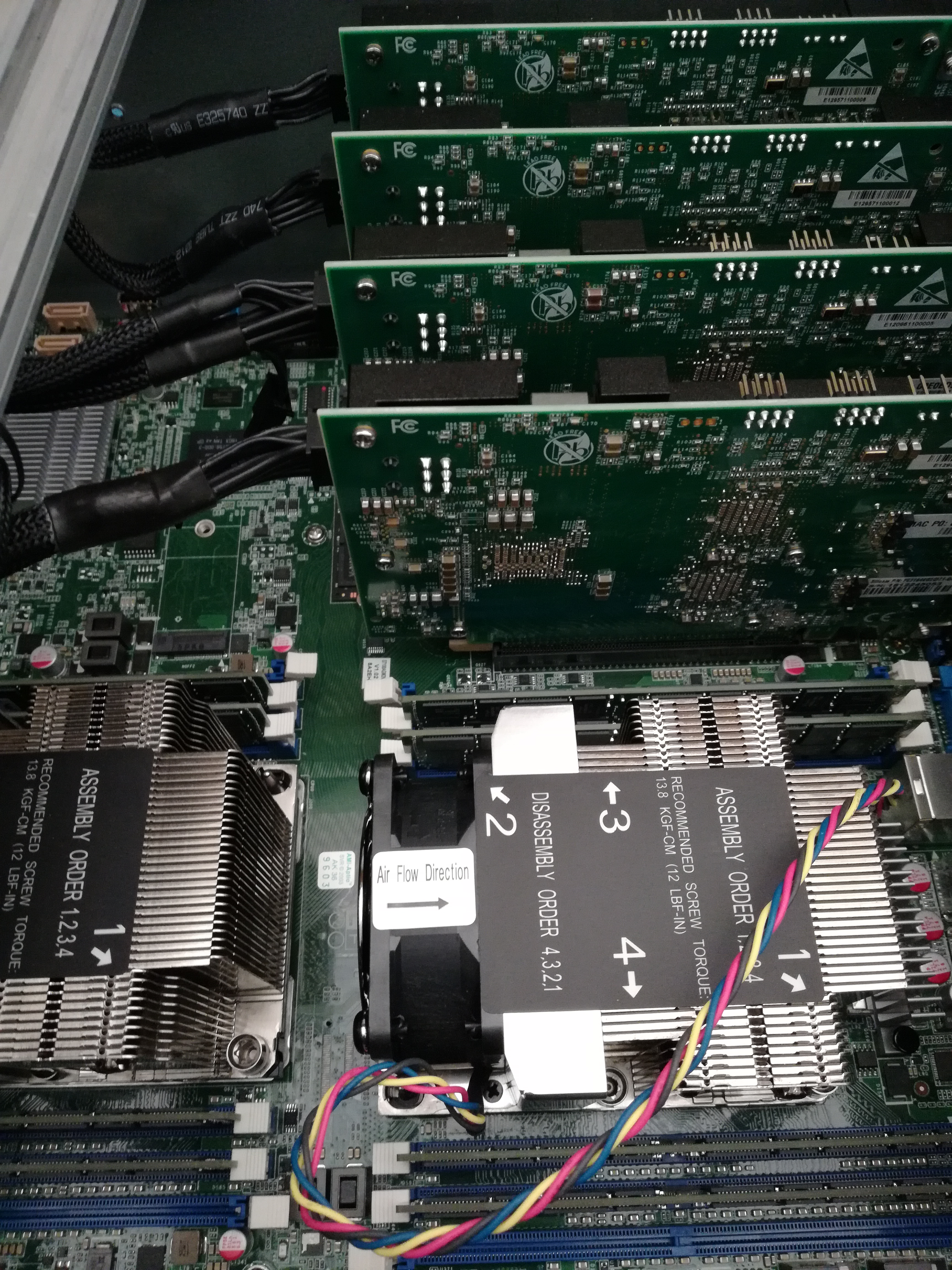}
    \caption{\textbf{Interior of the baseband recorder backend.}The baseband recorder architecture features four server grade network cards connected via a PCIeX16 slot to two CPU sockets, each of which can access 512 gigabytes (GB) of RAM with low latency. While awaiting a dump trigger from CHIME, our baseband recorder runs a custom version of the \texttt{kotekan} software framework which buffers 40 seconds of complex-valued baseband data for 256 of the Pathfinder F-engine's 1024 frequency channels. Four such baseband recorders could
    process the 0.8 Tb/s of data coming out of the Pathfinder, or an outrigger with similar data throughput. A full parts list is provided in Appendix~\ref{sec:parts}.}
    \label{fig:recorder}
\end{figure}

The CHIME Pathfinder was built prior to CHIME and is used for ongoing technology development for projects such as CHIME/FRB Outriggers. It has approximately one eighth of the collecting area of CHIME and operates on an independent clock. The effective baseline of Pathfinder is approximately 385.42 m due East, 50.43 due South, and 5.17 m lower than that of CHIME. It consists of two 20-m $\times$ 40-m cylinders which have the same field of view as CHIME, and have 64 dual-polarization antennas per
cylinder for a total of 256 correlator input channels. The Pathfinder shares the same F-engine architecture as CHIME, and runs on an independent GPS-disciplined crystal oscillator from that of CHIME. However, in contrast to a full FX correlator, the Pathfinder F-engine feeds baseband data to a baseband recorder backend. This backend, shown in Fig.~\ref{fig:recorder}, was built to demonstrate the technique of triggered VLBI observations for CHIME/FRB Outriggers. Using four server-grade network
cards which each provide $80$ Gb/s of bandwidth, the recorder stores baseband data in RAM for a quarter of CHIME/FRB's 1024 frequency channels, spaced approximately evenly across the band, at an input data rate of 204.8 gigabits per second (Gb/s) (for details, see Appendix~\ref{sec:parts}). Our ring buffer architecture is implemented in \texttt{kotekan}\footnote{\texttt{https://github.com/kotekan/kotekan}}, a flexible and efficient software framework written in C++ for real-time data processing for digital radio astronomy~\citep{KotekanTechnical}. 

\section{Interferometric Localization}\label{sec:localization}
\subsection{Detection at CHIME}\label{subsec:detectionchime}
CHIME/FRB features a real-time processing pipeline which coarsely estimates the DM, time of arrival, and signal-to-noise ratio of dispersed radio transients~\citep{FRBSystemOverview}. Upon detecting a sufficiently bright transient, a classification algorithm filters out false positives from radio frequency interference and known pulsars. Successful classification of a dispersed radio transient as an FRB triggers the dump of $\approx 100\,\text{ms}$ of baseband data to disk at both telescopes with subsecond latency.

Prior to data transfer and cross correlation, the baseband data from just the CHIME/FRB instrument are processed to estimate the FRB's dispersion measure and sky position. This is done by beamforming baseband data from CHIME/FRB's 2048 correlator inputs towards a grid of sky positions around the detection position, calculating the signal-to-noise ratio of the burst detection in each beam, and then fitting a 2D Gaussian model to the resulting intensity map of the signal. Finally, we perform
coherent dedispersion to the optimal dispersion measure maximizing the burst signal-to-noise ratio and form a tied-array beam to the refined coordinates provided by this so-called ``baseband pipeline''~\citep{michilli2020inprep}. From here on we denote the beamformed baseband data from CHIME as $F^C_{\nu b t}$. Here, $C$ stands for CHIME, while $\nu$ represents the frequency channel ($N_\nu = 1024$) ranging from 400-800 MHz. The integer $b$ is the ``beam number'', reflecting the fact that a single dump of full-array baseband
data can be beamformed to multiple sky positions in both polarizations (north-south and east-west, hereafter NS and EW); $b$ ranges from $1,2,\ldots,N_b$ where $N_b = 2 N_p$ and where $N_p$ is the number of unique sky positions. Finally, $t$ is the time index, measured in units of $\SI{2.56}{\micro\s}$. We calculate the flux as a function of frequency channel, polarization, and time block, albeit a lower time resolution indexed by $T$:
$$S^C_{\nu b T} = \sum_{t = T}^{t = T+t_{int}}|F^C_{\nu b t}|^2.$$
Setting the integration time $t_{int} = \SI{40.96}{\micro\s}$ yields the plots in Fig.~\ref{fig:chime-auto-waterfall}. 

\subsection{FRB Cross Correlation Pipeline}
Our cross correlation pipeline picks up where the baseband pipeline leaves off. Due to the reduced sensitivity of the Pathfinder, we only cross-correlate the baseband data from bright FRBs. We calculate beamformed baseband at both telescopes ($F^{C}_{\nu b t}$ and $F^{P}_{\nu b t}$), and divide the baseband data into segments of $\SI{40.96}{\micro\s}$. For each segment we calculate the complex temperature-normalized visibility $V^{CP}_{\nu b T}$ as a function of frequency,
polarization/beam, and time
block $T$ as we did previously for the flux. 
\begin{equation}
    V^{CP}_{\nu b T} = \dfrac{\sum_{t = T}^{t = T + t_{int}} F^C_{\nu b t} \overline{F^{P}_{\nu b t}}}{\sqrt{\sum_{t' = T}^{t' = T + t_{int}}  || F^C_{\nu b t'} ||^2 \sum_{t'' = T}^{t'' = T + t_{int}} || F^P_{\nu b t''} ||^2}}
    \label{eq:unweightedvis}
\end{equation} 

\begin{figure}[t]
    \centering
    \includegraphics[width = 0.485\textwidth,trim = 85 50 0 100,clip]{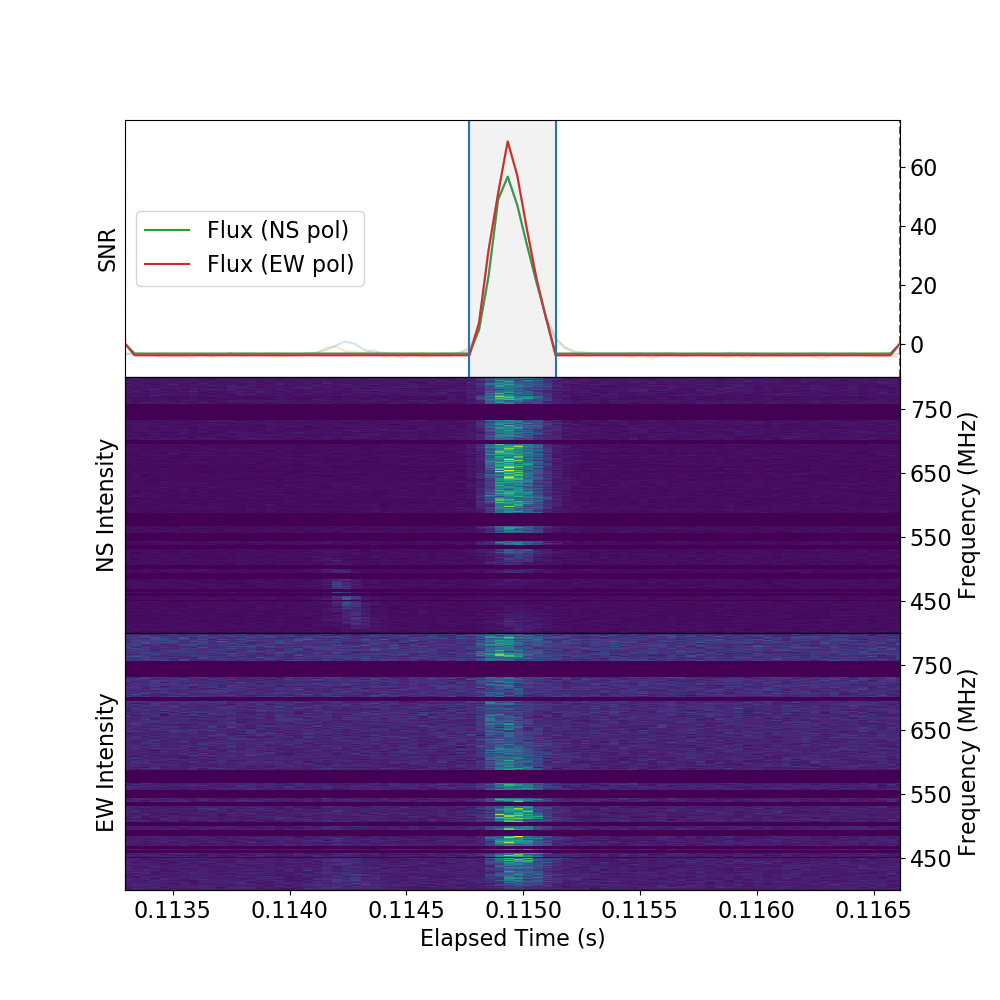}
    \caption{\textbf{CHIME waterfall plot for FRB 20191219F.} At UTC 2019-12-19T16:51:34, the detection of an FRB in CHIME triggered a simultaneous dump of channelized voltage data at CHIME/FRB and the CHIME Pathfinder. After nulling channels containing radio frequency interference, we beamform the baseband data at the optimum position calculated by the baseband pipeline, and plot the flux of the burst as a function of time and frequency in the 400-800 MHz band.}
    \label{fig:chime-auto-waterfall}
\end{figure}

\begin{figure}[t]
    \centering
    \includegraphics[width = 0.485\textwidth,trim = 0 50 0 100,clip]{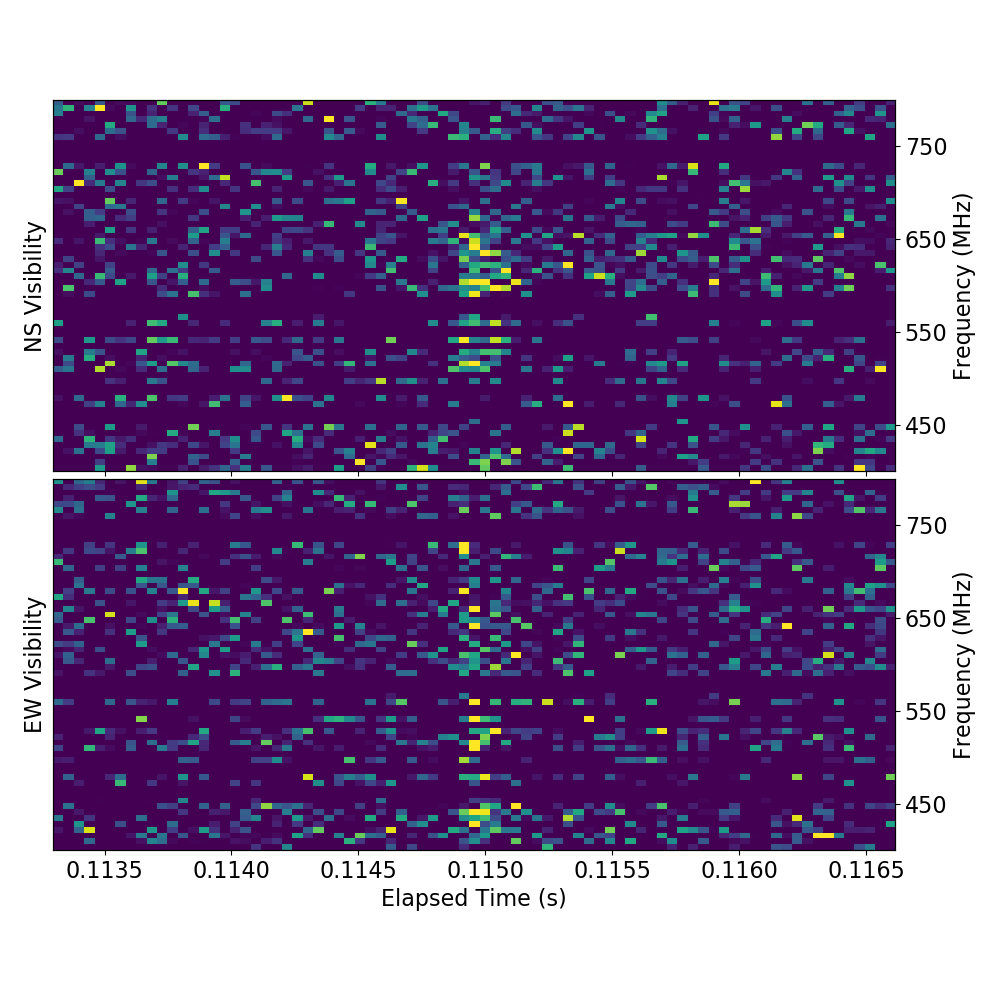}
    \caption{\textbf{Absolute magnitude of the temperature-normalized visibility between CHIME Pathfinder and CHIME/FRB, in both the north-south and east-west polarizations, calculated and as a function of time and frequency as in Eq.~\ref{eq:weightedvis}.} The morphology of the pulse as it appears in cross-correlation matches that detected at CHIME/FRB , revealing the detection of FRB 20191219F in cross-correlation between the two telescopes.}
    \label{fig:xcorr-waterfall}
\end{figure}

The quantity $V^{CP}_{\nu b T}$, like the baseband data, is complex-valued. For geometric delays shorter than $\SI{2.56}{\micro\s}$ the information about the geometric delay is completely encoded in the phase of the numerator of $V^{CP}_{\nu b T}$. The denominator ensures that increasing the system temperature (i.e. scaling any of the $F_{\nu b t}$ by a constant factor) does not affect $|V^{CP}_{\nu b T}|$. Hence, $|V^{CP}_{\nu b T}|$ as plotted in Fig.~\ref{fig:xcorr-waterfall} measures the strength of the
cross-correlation independently of the system temperature. The morphological similarity of $|V^{CP}_{\nu b T}|$ in Fig.~\ref{fig:xcorr-waterfall} and $S_{\nu b T}$ in Fig.~\ref{fig:chime-auto-waterfall} allows us to unambiguously interpret our cross-correlated baseband data as a genuine FRB detection. We cross-correlate the NS polarizations and EW polarizations at both telescopes separately; since the two telescopes' polarization axes differ by only $\approx 2$ degrees, this approach is close to optimal. 

While the above visibilities are sufficient for assessing a detection, for astrometric precision it necessary to minimize the uncertainty on the phase of the visibility. To do this, we formed a set of visibilities in which we integrated over the entire $\approx\SI{100}{\milli\s}$ baseband dump to reduce statistical uncertainty of the visibility phase. In addition, for the beams with pulsed emission, we perform the integration with the help of a real-valued time-domain matched filter, $h_t$, constructed from the pulse's intensity profile as detected in CHIME autocorrelation (i.e. the curves shown in the top panel of Fig.~\ref{fig:chime-auto-waterfall}).

\begin{equation}
    V^{CP}_{\nu b} = \dfrac{\sum_{t} F^C_{\nu b t} h_t \overline{F^{P}_{\nu b t}}}{\sqrt{\sum_{t'} || F^C_{\nu b t'} ||^2 \sum_{t''} || F^P_{\nu b t''} ||^2}}
    \label{eq:weightedvis}
\end{equation}
The filter is normalized to have $\langle h_t \rangle = 0$ and $\langle h_t^2 \rangle = 1$. The former constraint enables optimal rejection of steady sources of correlated voltage signals other than the pulse of interest, and the latter constraint ensures that the noise variance of the data is preserved.

\subsection{Synoptic Calibration Technique}\label{subsec:calib}

Our calibration technique fundamentally relies on in-field steady sources to keep the two telescope backends synchronized over the $\sim 10$ second duration of the dispersed burst.
Each array only needs to be individually synchronized once per day during the transit of a bright radio calibrator, to re-compensate for the slow thermal expansion of cables between the antennas and the correlator. However, since CHIME and the Pathfinder are each synchronized to independent ovenized crystal oscillator clocks, the time difference between the two arrays jitters on timescales of minutes. Clock jitter and differences in the telescopes' analog chains introduce an unknown instrumental phase between the two telescopes which must be calibrated near or during the time of observation. 

To solve for the instrumental phase, we used the fact that the primary beams of CHIME and Pathfinder completely overlap and that their large size virtually guarantees that there will be $\sim 5-10$ bright NVSS~\citep{condon1998nrao} calibrators ($S_{1.4GHz} > 1.5$\,Jy) detectable with a high signal-to-noise ratio in $\SI{100}{\milli\s}$ of integration time. For each observation, we selected seven of the brightest NVSS calibrators within 1.8 degrees from the local meridian. In total, we formed 16
beams from each triggered baseband dataset: one per polarization (north-south and east-west) per source (one transient and seven steady-source calibrators) towards catalogued positions of the calibrators as well as our initial estimate of the transient's position from the CHIME/FRB baseband pipeline. We calculated the visibility between the two telescopes as a function of beam and frequency as described in Eq.~\ref{eq:weightedvis} and we fit a delay model.

\subsection{Delay Model}
For each formed beam (indexed by $b$) and each frequency channel (indexed by $\nu$), our general delay model (more generally, a phase model) can be written as:

\begin{equation}
\Phi^{i}_{\nu b} = \phi^i_\nu + \vec{u}^i(t) \cdot\hat{n}_b + \dfrac{K\Delta DM(\hat{n}_b)}{\nu}
\label{eq:model}
\end{equation}
where $\phi^i_\nu$ is a free function representing the instrumental phase for the $i$th telescope, $\vec{u}^i(t)$ is the (time dependent) position of the $i$th telescope, $\hat{n}_b$ is the sky position of a source in the $b$th formed beam, and where the dispersive delay due to the ionosphere is a free function $\Delta DM(\hat{n}_b)$ and where the dispersion measure constant is taken to be $K = 1/(2.41\times10^{-4})$\,s\,MHz$^2$\,pc$^{-1}$\,cm$^3$. This simple model takes into account the
time-variable geometric delay and ionospheric delays; for simplicity we neglect small corrections such as tidal deformation that become necessary over long baselines. From here on, we suppress the time dependence of the telescope positions $\vec{u}^i(t)$. Also, since CHIME and Pathfinder are approximately co-located, the ionospheric delay only varies as a function of sky angle ($\hat{n}_b$) and not of position ($\vec{u}^i$).

While Eq.~\ref{eq:model} could in principle be fitted directly to the visibilities with a least-squares algorithm, in practice it is helpful to slow down, or ``fringestop'', the rapid phase variation of the visibility versus frequency to no more than a few radians over the telescope bandwidth using fiducial estimates for $\vec{u}^i$ and $\hat{n}_b$. This improves the robustness and convergence of the fit especially in the presence of noise. We denote these estimates with an additional subscript 0. First, we remove the geometric delay due to the nominal baseline ($\vec{u}_0^C - \vec{u}_0^P$), an estimate which is accurate to within a meter. We calculated the (uncalibrated) visibilities $V^{CP}_{\nu b}$, reducing our dataset to a set of $\sim 10^4$ complex numbers, one per frequency channel per formed beam. The phase of the uncalibrated visibilities after fringestopping can be modeled as
\begin{equation}
\label{eq:uncalib}
\phi^{CP}_{\nu b} = \Phi^{C}_{\nu b} - \Phi^{P}_{\nu b} = \phi^{CP}_\nu + (\vec{u}^C - \vec{u}^P) \cdot\hat{n}_b - (\vec{u}^C_{0} - \vec{u}^P_{0}) \cdot\hat{n}_{b,0}
\end{equation}
where $\phi^{CP}_\nu$ represents the differential instrumental phase between CHIME and Pathfinder, where $(\vec{u}^C - \vec{u}^P)$ is the true baseline, where $\hat{n}_b$ are the true positions, and where the last term encodes our fringestopping using nominal estimates of the sky positions and baseline. Note that the ionosphere term in Eq.~\ref{eq:model} is identical for each telescope and does not appear in Eq.~\ref{eq:uncalib}.
Since the differential instrumental phase is independent of sky pointing, we designate two reference beams ($B$) to use as phase references for the NS and EW polarizations of the telescope. We remove the differential instrumental phase by calculating $ \mathcal{V}_{\nu b} \equiv V_{\nu b} / V_{\nu B}$. We define $\sigma_{\nu b}$ to be the uncertainty on $\mathcal{V}_{\nu b}$, and denote the amplitude and phase of $\mathcal{V}_{\nu b}$ as $A_{\nu b}$ and $\varphi_{\nu b} \equiv \phi_{\nu b} - \phi_{\nu B}$ respectively.
\subsection{Fringe Fitting}~\label{subsec:mle}
After applying this calibration procedure, the phase of the fringestopped and calibrated visibilities which we fit to our delay model is
\begin{equation}
\label{eq:calib}
\varphi^{CP}_{\nu b} = (\vec{u}^C - \vec{u}^P) \cdot (\hat{n}_b - \hat{n}_B) - (\vec{u}^{C}_0 - \vec{u}^{P}_0) \cdot (\hat{n}_{b,0} -\hat{n}_{B,0})
\end{equation}

\begin{figure}
    \centering
    \includegraphics[trim = 10 0 0 0 ,clip,width=0.485\textwidth]{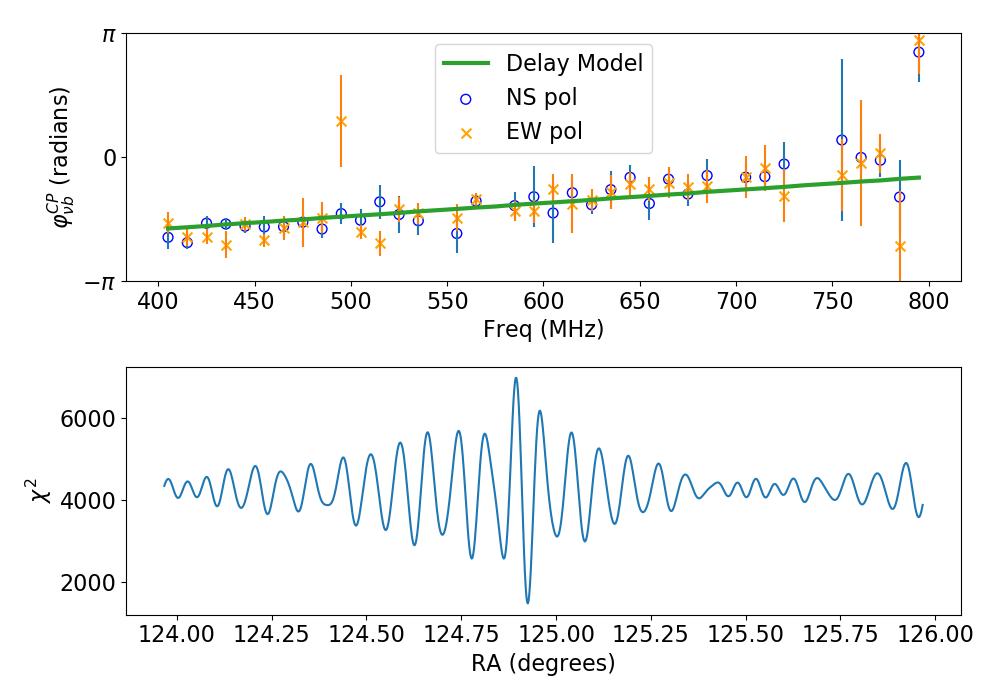}
    \caption{\textbf{Top: Successful fringe fit for FRB 20191021A}. We plot the slowly-varying phase $\varphi^{CP}_{b \nu}$ of the CHIME--Pathfinder visibility as a function of frequency in the NS and EW polarizations. To guide the eye, we bin over frequency channels with a resolution of $16$ MHz, and overlay the corresponding best-fit delay model (solid line). \textbf{Bottom:  Maximum likelihood $\chi^2$ statistic as a function of RA.} The log-likelihood function (negative of Eq.~\ref{eq:logl}) shows a clear minimum at the best-fit position of the FRB. Though we are fitting $N \approx 512$ visibilities, systematic effects such as a differential beam phase and confused calibrators prevent the $\chi^2$ statistic from reaching its expected value of $\approx 512$ at its minimum in parameter space. In addition, we slightly underestimate the thermal noise on the visibility, not taking into account the increased system temperature when the transient is on.}
    \label{fig:phase-vs-freq}
\end{figure}

\begin{figure}
    \centering
    \includegraphics[width = 0.485\textwidth]{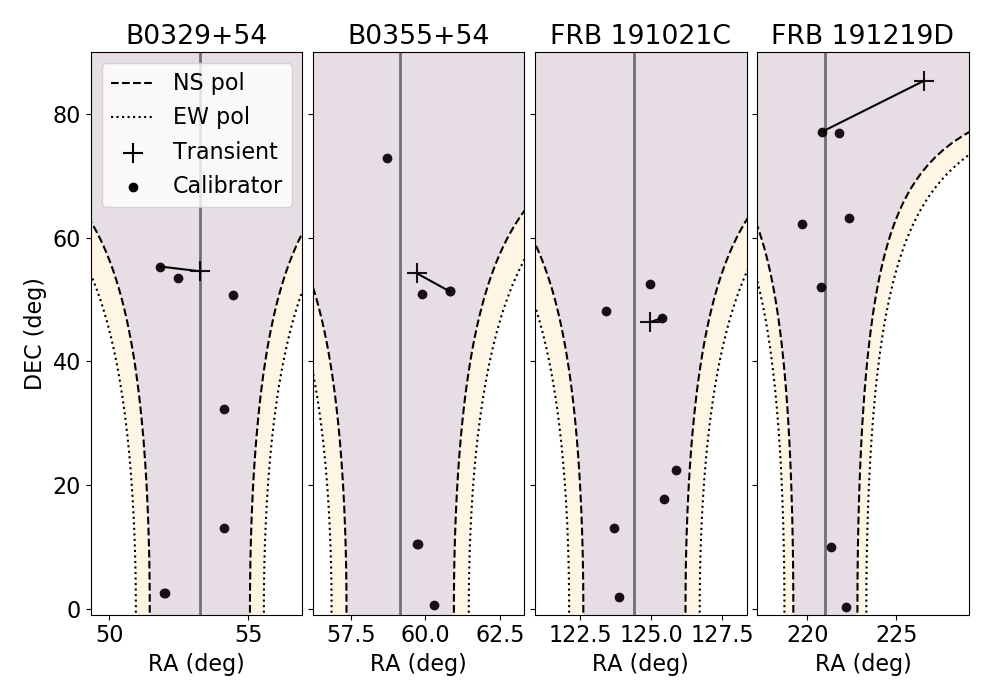}
    \caption{\textbf{Sky maps of the four fields we observed}, with a `$+$' denoting the approximate position of the pulsar/FRB, and bright NVSS calibrators with $S_{1.4 GHz} > 1.5$ Jy  indicated with black dots. The thick black lines denote the calibrator used to phase-reference each pulsar/FRB. Contours denote the FWHM of the primary beam of both telescopes ~\citep{CHIMECalibration} in the NS and EW polarizations at 600 MHz. The vertical black bars denote the meridian at the time of observation.}
    \label{fig:fields}
\end{figure}

With a good guess of the baseline offset, Eq.~\ref{eq:calib} varies slowly as a function of frequency and can be fitted to extract sky localizations and baseline information, as shown in Fig.~\ref{fig:phase-vs-freq}. First, using $\sim 10$ auxiliary $\SI{100}{\milli\s}$ snapshots similar to those shown in Fig.~\ref{fig:fields}, each targeting $\approx 7$ sufficiently-bright NVSS calibrators (for which $\hat{n}_b = \hat{n}_{b,0}$) at a wide range of sky positions, we determine the remaining baseline offset $\delta \vec{u} \equiv (\vec{u}^C - \vec{u}^P) - (\vec{u}_0^C - \vec{u}_0^P)$. Next, fixing $\delta \vec{u}$, we can determine the unknown sources' offsets from their nominal positions, denoted by $\delta \hat{n_b} \equiv \hat{n}_b - \hat{n}_{b,0}$. Note that our approximately east-west baseline make us insensitive to the declination of sources in the sky, and that the sky positions of sources we are observing (all close to the local meridian) make our data insensitive to east-west baseline errors.

The parameters $\delta \vec{u}$ and $\delta \hat{n}_b$ are estimated by maximizing the likelihood $\mathcal{L}$ using an expression that does not depend on the intrinsic emission spectra of any of the sources. Since only the phase of the visibility is sensitive to astrometric quantities, we can analytically marginalize over the amplitude $A_{\nu b}$ of the calibrated visibilities without losing phase information. We suppress the superscript in Eq.~\ref{eq:calib}, treating it as a free function $\varphi_{b \nu}$ of sky positions and baseline parameters which we collectively refer to as $\lambda$.
Assuming a uniform prior and applying Bayes's theorem we can write the posterior distribution of $\lambda$ with a $\chi^2$ maximum likelihood estimator. Integrating over the amplitude of the visibility $A_{\nu b}$ simplifies our full $\chi^2$ likelihood to its form in Eq.~\ref{eq:logl}.
\begin{align}
P(\lambda | \mathcal{V}_{\nu b}) \propto &P(\mathcal{V}_{\nu b} | \lambda)\nonumber \\ 
\propto &\exp\left(-\dfrac{1}{2}\sum_{\nu b}\dfrac{||\mathcal{V}_{\nu b} - A_{\nu b}\exp(i\varphi_{\nu b}(\lambda))||^2}{\sigma^2_{\nu b}}\right)\nonumber\\
\propto &\exp\left(-\dfrac{1}{2}\sum_{\nu b} \text{Im}[\mathcal{V}_{\nu b} \exp(-i\varphi_{\nu b}(\lambda)) / \sigma_{\nu b}]^2\right).\nonumber\\
\log \mathcal{L} \propto & -\dfrac{1}{2}\sum_{\nu,b} \text{Im}[\mathcal{V}_{\nu b} \exp(-i\varphi_{\nu b}(\lambda)) / \sigma_{\nu b}]^2.\label{eq:logl}
\end{align}

Intuitively, this can be understood as follows. If the delay model allows us to perfectly derotate the $\mathcal{V}_{\nu b}$ to the real axis of the complex plane, the imaginary part of $\mathcal{V}_{\nu b}$, normalized by its standard deviation, will be minimized and will be a zero-mean, unit-variance Gaussian random variable. Hence, the sum of squares follows a $\chi^2$ distribution with $N_b \times N_\nu$ degrees of freedom, and minimizing $\chi^2$ allows us to recover the best fit parameters $\lambda$ without ever explicitly fitting any spectra. Our FRB localizations are summarized in Table ~\ref{tab:localizations}.

Statistical uncertainties are estimated by jack-knifing our data over frequencies: we can divide our calibrated visibilities $\mathcal{V}_{\nu b}$ into 9 different ``frequency combs'', spaced evenly across our band. By leaving out one comb at a time and repeating our $\chi^2$ analysis, we can inspect the resulting likelihood curves and reject frequency-local RFI, which would show up as a discrepancy between different jack-knifed realizations of our analysis. We estimate the statistical error on our localizations using our jack-knifed samples in accordance with~\cite{mcintosh2016jackknife}.

\begin{table*}[htbp]
    \centering
    \caption{\textbf{Localization of Known Pulsars and Fast Radio Bursts Detected by CHIME/FRB.} We report the DM, nominal sky position, and observing epoch during which we collected baseband data on each source. For pulsars, the nominal RA and DEC (in degrees) are taken from the ATNF catalog~\citep{manchester2005australia}. For FRBs, we instead report the nominal RA and DEC at which the FRB was detected by CHIME/FRB's real-time pipeline. We report the measured RA from our localization pipeline with statistical uncertainties and systematic offset of each source from its true position. For the pulsars, the systematic offset is known, and for the FRBs, the systematic offsets are extrapolated from those of pulsars (see text and Fig.~\ref{fig:sys}). We are unable to unambiguously identify a single host galaxy with our current localization precision.}
    \begin{tabular}{|lrrrrrr|}
    \hline
    Source & DM & RA (nominal) & DEC (nominal) & Epoch (MJD) & RA (measured) $\pm$ Stat & Offset (deg)\\
    \hline
PSR B0329+54 & 26.776  &  53.24770 & 54.57860 & 58772.412  & $53.24538 \pm 0.00017$ & $-0.00232$ \\
PSR B0329+54 &         &           &          & 59032.701  & $53.25361 \pm 0.00029$ & $0.00591$ \\
PSR B0329+54 &         &           &          & 59034.696  & $53.25339 \pm 0.00021$ & $0.00568$ \\
PSR B0355+54 & 57.142  &  59.72391 &  54.2205 & 59033.713  & $59.72725 \pm 0.00101$ & $0.00334$ \\
FRB 20191021A  & 388.659 & $124.92$  & $46.39$   & 58777.595  & $124.92521 \pm 0.00044$ & $\pm\sim 0.005$\\
FRB 20191219F & 464.560  & $225.92$ & $85.44$ & 58836.702  & $226.56408 \pm 0.00694$ & $\pm\sim 0.05$ \\
    \hline
    \end{tabular}

    \label{tab:localizations}
\end{table*}

\subsection{Systematic Errors}\label{subsec:sys}
Over short baselines, most radio sources remain unresolved and there is no shortage of calibration sources in the sky. While our database of NVSS sources serves as an abundant calibrator network, it also means that the probability of having two sources within a formed beam (FWHM $\sim 0.3\,\text{deg}^2$) is non-negligible. While we filtered out bright calibrator candidates that are too close to each other, we are forced to assume that the remaining calibrators are true point sources. For example, we cannot eliminate the possibility that the emission of sources at low frequencies (400-800 MHz) is offset from the catalogued survey position at $\SI{1.4}{\giga\hertz}$. Angular offsets of the calibrator's emission region from its catalog position would severely impact our measurements in two major ways.\\
First, astrometric discrepancies for sources used as a delay center could directly lead to localization errors, creating a systematic offset in the measured RA. This effect should be smaller than the formed beam size and should be independent of the calibrator's angular distance from the transient of interest.

The more serious impact of astrometric discrepancies is on baseline determination. Eq.~\ref{eq:calib} implies that an inaccurate determination of the baseline translates to a systematic localization offset proportional to the on-sky distance between the target of interest ($\hat{n}_b$) and the delay center ($\hat{n}_{B}$).

To quantify the systematic offsets in our RA measurements, we conducted triggered observations of pulsars, which are also summarized in Table~\ref{tab:localizations}. We added rules to the event classifier in the real time FRB detection pipeline to allow bright pulses from known pulsars to trigger a baseband dump, in the same way that an FRB would. In this way, we collected baseband data for three bright single pulses from PSR B0329+54 and one from PSR B0355+54, and localized the pulsars as if they were FRBs. We estimated the systematic errors in our localization analysis using the discrepancy between our results and the pulsars' known position, corrected for their proper motion.

\begin{figure}
    \centering
    \includegraphics[width = 0.485\textwidth]{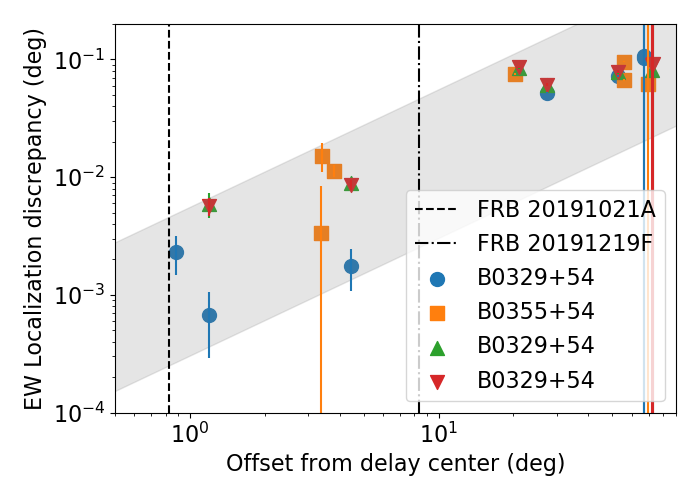}
    \caption{\textbf{Deviation of the localized positions of B0329+54 and B0355+54 from their true positions along the RA direction as calculated by using different NVSS calibrators as delay centers.} The discrepancy in degrees is quantified as the coordinate offset $\Delta \text{RA} \times \cos(\text{DEC})$ and is plotted with $3\sigma$ statistical error bars. We compute localizations for the same pulsar using different phase centers to study the effect of using different delay centers on the same transient. The shaded gray band is drawn to guide the eye and allows us to estimate the systematic localization offset of the two FRBs, whose closest calibrators are 0.8 and 8 degrees away respectively.}
    \label{fig:sys}
\end{figure}

We phase reference the pulsar position to the 7 in-beam NVSS calibrators, whose sky positions are as far as 60 degrees away from the pulsar. We plot the astrometric localization error against the angular distance between the pulsar and the delay center in Fig.~\ref{fig:sys}. We find that the astrometric discrepancy is roughly linearly proportional to the on-sky distance to the calibrator, and that using the nearest on-sky calibrator minimizes discrepancies from the catalogued positions of pulsars even with truly simultaneous phased-array observations through the same ionosphere. We attribute this discrepancy chiefly to a static baseline determination error corresponding to time delays of less than a nanosecond. To estimate the magnitude of systematic uncertainty in our FRB localizations, we find the intersection of the upper edge of the shaded area in Fig.~\ref{fig:sys} with the on-sky distance to the nearest calibrator to each FRB. 

In addition to an unknown static baseline error, the effective phase center of a beamforming telescope drifts slightly every day. The effective phase center position is the centroid of active antenna positions weighted by their sensitivity, and the centroid drifts from day to day on the order of $\sim \text{cm}$ because a slightly different set of antennas are flagged (i.e. nulled) every day due to factors like rain causing increased noise in certain antennas. We take this effect into account during tied-array beamforming, but the current baseline positions are not yet constrained at a level to measure this day-to-day drift in astronomical data. Using a larger sample of pulsars at a wide range of declinations for baseline determination, not just validation, will reduce our systematic error floor and improve our ability to phase reference our observations to calibrators far away on the sky.
\section{Discussion and Conclusion}
We have developed baseband recording hardware and software capable of handling the high data rate of wideband, multi-element radio interferometers such as CHIME for VLBI observations (Section~\ref{sec:instrumentation}). Also, we have demonstrated a calibration technique that exploits CHIME's wide field of view to localize several radio transients detected by CHIME/FRB and the CHIME Pathfinder in the same spirit as VLBI (Section~\ref{sec:localization}). In an automatically triggered $\approx \SI{100}{\milli\s}$ duration baseband capture at CHIME and Pathfinder, we can simultaneously detect a single FRB in cross correlation between CHIME and Pathfinder, as well as multiple calibrators for phase referencing our telescopes. 

We have developed efficient maximum likelihood estimators to perform fringe fitting in the absence of knowledge about the FRB spectrum(Section~\ref{subsec:mle}), and have localized FRB 20191021A and FRB 20191219F with statistical uncertainties of $1.6$ and $25$ arcseconds respectively along one direction in the sky (Table~\ref{tab:localizations}). Using single pulses from bright pulsars we have characterized the systematic errors on our FRB localizations (18 arcseconds and 3 arcminutes respectively) which are dominated by errors in baseline determination using NVSS calibrators (Section~\ref{subsec:sys}). 

Eventually, CHIME/FRB Outriggers will include stations at baselines of thousands of kilometers to achieve an astrometric precision of $\sim 50$ milliarcseconds. This precision is roughly matched to that of the best optical telescopes, and will allow for detailed followup studies of FRB host environments within their host galaxies. To achieve our goal, we anticipate a very different set of challenges from those presented here. Over long baselines, the ionospheric phase shift can vary by as much as $\Delta \text{DM} \sim 10^{-5}$ (corresponding to a time delay of $\sim\SI{200}{\nano\s}$ as a function of sky position at sub-gigahertz frequencies). Achieving high astrometric precision will require removing this effect with observations of calibration sources close to the FRB on the sky. The relatively uncharted territory of low-frequency VLBI calibrators poses a major challenge for scaling CHIME/FRB VLBI observations to continental baselines.

One option is to use bright pulsars for phase referencing observations with CHIME/FRB Outriggers, especially for hour angles close to the Galactic plane where pulsars are most abundant. Pulsars have the advantage of being transient and compact point sources, helping to eliminate confusion noise and the effect of uncertain calibrator morphology on our astrometric precision. Though the astrometric positions of some pulsars are known at the $10-20$ milliarcsecond level, including less precisely localized pulsars in the calibrator network of CHIME/FRB Outriggers will improve astrometric localizations of those pulsars as observations accumulate over time.

For hour angles where pulsars are sparse, phase referencing after the real-time detection of an FRB can be done by using a dense network of steady-source VLBI calibrators all over the northern sky, particularly near the celestial pole in the constant-coverage area of CHIME's primary beam.

Following pioneering low-frequency VLBI surveys by~\citet{garrett2005deep} and ~\citet{lenc2008deep}, the advent of the International LOFAR Telescope has made systematic surveys of the low-frequency sky possible. The LOFAR Snapshot Calibrator Survey~\citep{moldon2015lofar} has demonstrated that high quality, compact VLBI calibrators at low frequencies tend to be bright at 328 MHz ($S = 0.1-1$~Jy) and have a flat low-frequency spectrum. Recent results from the ongoing LOFAR Long-Baseline Calibrator Survey~\citep[LBCS,][]{jackson2016lbcs} project the density of high-quality VLBI calibrators over long baselines to be $\sim 1\text{ deg}^{-2}$. While the LBCS covers even lower frequencies than those relevant for CHIME/FRB Outriggers, an understanding of promising low-frequency calibrators on long baselines will be crucial for future VLBI observing campaigns with CHIME/FRB Outriggers. The instrumentation and analysis techniques developed in this paper, combined with a dense network of pulsars or compact low-frequency VLBI calibrators, will pave the way for transformative studies of FRB host environments and of the intergalactic medium over long baselines with CHIME/FRB Outriggers.



\software{\texttt{numpy} \citep{numpy}, \texttt{scipy} \citep{virtanen2020scipy}, \texttt{matplotlib}~\citep{hunter2007matplotlib}}

\acknowledgments
{We thank the CHIME Collaboration for use of the Pathfinder, and the staff at the Dominion Radio Astrophysical Observatory and Ev Sheehan for their hospitality and efforts to ensure the smooth deployment of our instrumentation. C. L. was supported by the U.S. Department of Defense (DoD) through the National Defense Science \& Engineering Graduate (NDSEG) Fellowship. M.B. is supported by an FRQNT Doctoral Research Award. This research is funded in part by the Gordon and Betty Moore
Foundation and the NEC Corporation Fund for Research in Computers and Communication. FRB research at UBC is supported by an NSERC Discovery Grant and by the Canadian Institute for Advanced Research. V.M.K. holds a Distinguished James McGill Chair and the Lorne Trottier Chair in Astrophysics \& Cosmology and receives support from an NSERC Discovery Grant and Herzberg Award, from an R. Howard Webster Foundation Fellowship from the Canadian Institute for Advanced Research (CIFAR), and from the FRQNT Centre de Recherche en Astrophysique du Quebec. D. M. was supported by the Banting Postdoctoral Fellowships Program. P.S. is a Dunlap Fellow and an NSERC Postdoctoral Fellow. The Dunlap Institute is funded through an endowment established by the David Dunlap family and the University of Toronto. The CHIME/FRB baseband recording system was funded in part by a CFI John R. Evans Leaders Fund award to I.H.S.}
\appendix

\section{Baseband Recorder Parts List}\label{sec:parts}
Our recorder uses 1 terabyte of RAM to buffer approximately 40 seconds of baseband data corresponding to dispersion measures of up to $\approx 2000$ pc / cm$^3$ upon receiving a trigger from CHIME/FRB's real-time detection pipeline. A photograph of the inside of the node is shown in Fig.~\ref{fig:recorder}, and a full parts list is given in Table~\ref{tab:parts-list}. Future recorders may feature an auxiliary buffer or GPUs for real-time beamforming capabilities~\citep{CHIMEBeamforming}, which will facilitate longer integration times on fainter calibrators, though this technical capability is not necessary for our bright calibrators.
\begin{table*}[thb]
    \centering
    \caption{\textbf{Components used in the prototype baseband recorder for CHIME/FRB Outriggers.} The total cost of the recorder was less than $\$20$k USD in Spring 2019 and was dominated by the cost of the high-density RAM.}
    \begin{tabular}{|l|r|r|}
    \hline
    Parts & Part Number & Specifications (each)\\
    \hline
    Motherboard & $1\times$ TYAN Tempest EX S7100-EX & 4$\times$ PCIeX16, 3$\times$ PCIeX8, 2 sockets \\
    CPU     & 2$\times$ Intel Xeon Silver 4116 & 12 cores (hyperthreaded) $\times$ 2.10 GHz \\
    RAM     & 8$\times$ HYNIX HMAA8GR7A2R4N-VN & 128 GB \\
    Network & 4$\times$ Silicom PE 31640G2QI71/QX4 & $2\times 4\times$10GbE \\
    \hline 
    \end{tabular}

    \label{tab:parts-list}
\end{table*}

\bibliographystyle{aasjournal}
\bibliography{radio,chimefrbpapers}

\end{document}